\documentclass{article}
\usepackage{amsmath,amssymb}
\usepackage{graphicx}%
\usepackage{verbatim}
\usepackage[T1,T2A]{fontenc}
\usepackage[cp1251]{inputenc}
\usepackage[english]{babel}
\newtheorem{thm}{Statement}
\def\stackunder#1#2{\mathrel{\mathop{#2}\limits_{#1}}}
\def\stackunder#1#2{\mathrel{\mathop{#2}\limits_{#1}}}

\newcommand{\Req}[1]{(\ref{#1})}

\newcommand{\Eps}{{\cal E}}

\begin{document}

\begin{center}
{\bf \Large The Statistical Model with Interpartial Scalar Conformally Invariant Interaction}. \\[12pt]
Yurii Ignat'ev\\
N.I. Lobachevsky Institute of Mathematics and Mechanics, Kazan Federal University, \\ Kremleovskaya str., 35, Kazan, 420008, Russia
\end{center}

\begin{abstract}
A closed mathematical model of the statistical self-gravitating system of scalar charged particles for conformal invariant scalar interactions is constructed on the basis of relativistic kinetics and gravitation theory. Asymptotic properties of the model are investigated in the ultrarelativistic limit. It is shown, that scalar charge density automatically
generates scalar field effective mass and the value of this mass is found. In the paper it is proved the asymptotic conformal invariance of constitutive equations in case of homogenous isotropic Universe. Also it is proved the asymptotic conformal invariance of field equations at the early stages of cosmological evolution.

{\bf keywords}{\it Gravitation, Relativistic Kinetics,
Scalar Interaction, Conformally Invariant, Thermodynamic Equilibrium.}

\end{abstract}

This work was founded by the subsidy allocated to Kazan Federal University for the state assignment in the sphere of scientific activities.

\section{~Introduction}
\label{intro}

In the early 1980-s the fundamentals of the relativistic kinetic theory of the statistical systems with scalar interaction were formulated \cite{Ignatev1,Ignatev2,Ignatev3,Ignatev4,kuza}. In these years the kinetic theory of particles with scalar interaction seemed to be conceptual theoretical construction with the only purpose of the kinetic theory completeness. However, since the discovery of dark matter factor in cosmology and Higgs bosons in the last decade the development of this theory becomes actual and required for the development of the theoretical physics.
In the one hand, the relativistic kinetic theory is a bridge connecting microscopic and macroscopic levels of matter description and on the other hand it is sufficiently rigid theoretical  structure  significantly reducing a possibility of speculative constructions. In the series of recent Author's works it was formulated the strict relativistic kinetic theory of the statistical systems of scalar charged particles based on the canonical microscopic dynamics and following accurate procedures of the macroscopic averaging; this extends the theory to the case of fantom scalar fields and negative effective particle masses \cite{YuNewScalar1,YuNewScalar2,YuNewScalar3,YuNewScalar4,YuNewScalar5,Yu_stfi14,Yu_stfi15,YuNewScalar6,Ignatev_print1,Ignatev_print2}\footnote{see
also monographes \cite{Yubook1,Yubook2}.}. Moreover, there were constructed the cosmological models based on such systems \cite{Mif,Yu_Sasha,Yu_Sasha_Misha_Mitya_stfi,Yu_Sasha_Misha_Mitya,Yu_Misha}. In these recent works the models with conformal non-invariant scalar field were considered. Let us notice that the first work on the kinetic theory of the statistical systems with scalar interaction \cite{Ignatev1} set the problem of asymptotic conformal invariance of the kinetic theory in the ultrarelativistic limit, however it did not provide a well defined answer on this question. The importance of this problem's solution for the contemporary cosmology and the degree of modern development of the kinetic theory of scalar systems with interparticle scalar interaction drove the research presented in this paper. When referring to invariant scalar fields we will imply the scalar fields which equations contain a
 ``conformal member'' $-R/6$.
%
\section{~The Strict Macroscopic Relations of The Relativistic Kinetic Theory}
In this chapter we write out the condensed strict relations of the relativistic kinetic theory not depending on the transformational properties of the scalar field obtained in the papers cited above.

\subsection{The Canonical Equations Of Motion Of Particles In The Scalar Field}
The canonical equations of motion of the relativistic particle in phase space $\Gamma$ have the following form (see e.g. \cite{Ignatev2}):

where $H_a(x,p)$ is a relativistically invariant Hamilton function of
the $a$-sort particle with a scalar charge $q^{(r)}_a$ in scalar field $\Phi_r(x)$:

\begin{equation}\label{H,m}
H(x,p)=\frac{1}{2} \left[m_*^{-1}(x)(p,p)-m_*\right]=0,
\end{equation}
\begin{equation}\label{m_*}
m_*=\sum\limits_r q^{(r)}_a\Phi_r
\end{equation}
is an effective mass of a particle,  $u^i=dx^i/ds$ is a particle's velocity vector.
The full derivative of the dynamic variables function $\Psi (x^{i} ,p_{k} )$ with an account of (\ref{Eq1}) can be represented in form:

\begin{equation} \label{Eq2}
\frac{d\Psi }{ds} =[H,\Psi],
\end{equation}
where there are introduced the invariant Poisson brackets:
\begin{eqnarray}\label{Eq3}
[H,\Psi ]=\frac{\partial H}{\partial p_{i}} \frac{\partial \Psi }{\partial x^{i} } -\frac{\partial H}{\partial x^{i} } \frac{\partial \Psi }{\partial p_{i} } ;\\
\label{HPsi}
\equiv \frac{1}{m_*}p^i\widetilde{\nabla}_i\Psi+\partial_i m_*\frac{\partial \Psi}{\partial p_i},
\end{eqnarray}
where $\widetilde{\nabla}_i$ {\it is an operator of covariant Cartan differentiation
}\footnote{Covariant derivative in the stratificagion $\Gamma$ \cite{Cartan}.} (see e.g.
 \cite{Bogolyub})\footnote{Covariant Cartan derivatives were first introduced in the
 relativistic statistics by A.A.Vlasov \cite{Vlasov}.}:
\begin{equation}\label{Cartan}%
\widetilde{\nabla}_i = \nabla_i +
\Gamma_{ij}^k p_k\frac{\partial}{\partial p_j},
\end{equation}
where $\nabla_i$ is an operator of covariant Ricci differentiation and
 $\Gamma^k_{ij}$ are Christoffel symbols of the second kind relative to metrics
 $g_{ij}$ of $X$ base.
Operator $\widetilde{\nabla}$ is defined so that:
\begin{equation}\label{9.11}%
\widetilde{\nabla}_ip_k \equiv 0
\end{equation}
%

\subsection{The Relativistic Kinetic Equations}
Let the following reactions run in plasma:
\begin{equation} \label{GrindEQ__50_}
\sum _{A=1}^{m} n _{A} a_{A} {\rm \rightleftarrows }\sum _{B=1}^{m'} n '_{B} a'_{B} ,
\end{equation}
where $a_{A} $ are particle symbols and $n _{A} $ are their numbers in each reaction channel.
As a result of the local correspondence principle
and the suggestion of 4d-point particle collisions in each act
of interparticle interaction the generalized momentum of interacting particles is conserved:
\begin{equation} \label{GrindEQ__49_}
\sum _{I} p_{i} =\sum _{F} p'_{i} ,
\end{equation}
where the summation is carried out by all initial ($I$), $p_{i} $ and
final ($F$), $p'_{i} $ states. Thus, the generalized momentums of the initial and final states are equal:
\begin{equation}\label{GrindEQ__51_}
p_I=\sum\limits_{A=1}^m\sum\limits_\alpha^{n_A} p^\alpha_A,
\quad p_F=\sum\limits_{B=1}^{m'}\sum\limits_{\alpha'}^{n'_B} p'\
\!\!^{\alpha'}_B.
\end{equation}
Distribution functions of particles are defined through the invariant kinetic equations \cite{Ignatev3}:
\begin{equation} \label{GrindEQ__52_}
m_*[H_a,f_a]=\mathrm{I}_a(x,p),
\end{equation}
where $\mathrm{I}_{a} (x,p_{a} )$ is an integral of collisions:
\begin{eqnarray} \label{GrindEQ__53_}
\mathrm{I}_{a} (x,p_{a} )= -\sum  n _{A} \int  '_{a}\delta ^{4} (p_{F} -p_{I} )& \nonumber\\
\times W_{IF} (Z_{IF} -Z_{FI} )\prod _{I,F} 'dP;&
\end{eqnarray}
where
$$W_{FI} =(2\pi )^{4} |M_{IF} |^{2}\cdot 2^{-(\sum  n _{A} +\sum  n '_{b}) } $$
is a scattering matrix of channel of reactions \eqref{GrindEQ__50_}, $|M_{IF}|$ are invariant
amplitudes from the initial state $I$ to final $F$;
\begin{eqnarray}\nonumber
Z_{IF} =\prod _{I} f(p_{A}^{\alpha } )\prod _{F} [1\pm f(p_{B}^{\alpha '} )];\\
Z_{FI} =\prod _{I} [1\pm f(p_{A}^{\alpha } )]\prod _{F}f(p_{B}^{\alpha '} ),\nonumber
\end{eqnarray}initial state
sign ``+'' corresponds to bosons while `-'' sign corresponds to fermions (see details in \cite{Ignatev3,Ignatev4}).
%
\subsection{The Transport Equations of The Dynamic Quantities}
The transport equations of the dynamic quantities are the strict integral-differential consequences of the relativistic kinetic equations with the assumption that 4-vector of generalized momentum
(\ref{GrindEQ__51_}) is conserved in all channels of elementary particles interaction:
\begin{equation}\nonumber
\begin{array}{l}
\displaystyle\nabla _{i} \sum \limits_{a} \int\limits _{P_0} \Psi _{a} f_{a}
p^i dP_{a} -\sum _{a} \int\limits _{P_0} f_{a} m_*[H_{a} ,\Psi _{a} ]dP_{a} \\[12pt]
\label{GrindEQ__54_}
=-\displaystyle\sum\limits _{by\; chanels} \int\biggl(\sum\limits _{A=1}^{m} n _{A} \Psi _{A} -
\sum\limits _{B=1}^{m'} n '_{B} \Psi '_{B} \biggr)\\[12pt]
\times\displaystyle\delta ^{4} (p_{F} -p_{I} )(Z_{IF} W_{IF} -Z_{FI} W_{FI} )\prod \limits_{I,F} dP,
\end{array} \nonumber
\end{equation}
where the summation is carried out by all reaction channels \eqref{GrindEQ__50_}.

At $\Psi _{a}=g_{a} $, where $g_{a} $ are certain fundamental charges conserved in reactions \eqref{GrindEQ__50_}, taking into account
\eqref{GrindEQ__49_}, \eqref{GrindEQ__51_} and \eqref{GrindEQ__54_}
we obtain transport equations for flux densities of plasma particles number:
\begin{equation} \label{GrindEQ__55_}
\nabla _{i} J_{G}^{i} =0,
\end{equation}
where:
\begin{equation} \label{GrindEQ__56_}
J_{G}^{i} =\sum _{a} \frac{2S+1}{(2\pi )^{3} } \; g_{a}
\int\limits_{P_0} f_a(x,p)p^{i} dP_0.
\end{equation}
is a density vector of the fundamental charge corresponding to charges $g_{a} $.

Assuming $\Psi _{a} =P^{k} $ in \eqref{GrindEQ__54_}, we obtain transport equations for plasma
energy-momentum:
\begin{equation} \label{GrindEQ__57_}
\nabla _{k} T_{p}^{ik} -\sum\limits_r\sigma_{(r)}\nabla ^{i} \Phi_{r} =0,
\end{equation}
where it is introduced the \textit{tensor of energy-momentum}
\begin{equation}\label{Tpl}
T^{ik}_p=\sum\limits_{a} \frac{2S+1}{(2\pi )^{3}}
\int\limits_{P_0} f_a(x,p)p^ip^kdP_0
\end{equation}
and \textit{scalar densities of plasma charge relative to scalar field $\Phi_r$}, $\sigma^{(r)}$:
\begin{equation}\label{sr}
\sigma^{(r)}=\sum\limits_a \sigma^{(r)}_a,
\end{equation}
where $\sigma^{(r)}_a$ -- \textit{ are scalar charge densities of $a$-component of plasma relative to scalar field $\Phi_r$}:
\begin{equation} \label{GrindEQ__58_}
\sigma^{(r)}_a =\frac{2S+1}{(2\pi )^{3} } m^*_a q^{(r)}_a
\int\limits_{P_0} f_a(x,p)dP_0,
\end{equation}

In particular, for charge singlet $(q,\Phi)$ conservation law \eqref{GrindEQ__57_} takes form:
\begin{equation} \label{GrindEQ__57_0_}
\nabla _{k} T_{p}^{ik} -\sigma \nabla ^{i} \Phi =0,
\end{equation}
where (see \cite{Ignatev3,Yu_stfi14}):
\begin{equation} \label{GrindEQ__59_}
\sigma =\Phi \frac{2S+1}{(2\pi )^{3} } q^{2}
\int\limits_{P_0)} f(x,p)dP_0.
\end{equation}
Let us notice that the forms of energy-momentum tensor (EMT) \eqref{Tpl} and scalar charge density
\eqref{GrindEQ__58_} found for scalar charge particles at given Hamilton function represent direct consequence of the canonical equations and the suggestion about total momentum conservation
 at local collisions of particles.
%
\subsection{Thermodynamic Equilibrium of Plasma in The Gravitational Field}
\label{I.III.1}Distribution functions in conditions of thermodynamic equilibrium due to so-called functional\\ Boltzmann equations take
 {\it locally-equilibrium} form:
\begin{equation}\label{3.1.9}
f^0_a (x, p_a) = \frac{1}{\displaystyle \mathrm{e}^{ - \nu_a +
(\xi, p_a)} \mp  1}\,,
\end{equation}
where upper sign corresponds to bosons, lower sign corresponds to fermions and vector $\xi^i(x)$ should be timelike:
\begin{equation}\label{3.1.10}
\xi^2 \equiv (\xi, \xi) > 0\,,
\end{equation}
and {\it reduced chemical potentials} $\nu_a$ should satisfy the series of conditions of
{\it chemical equilibrium}:
\begin{equation} \label{lambda_chem}
\sum _{A=1}^{m} n _{A} \nu_{A} =\sum _{B=1}^{m'} n '_{B} \nu'_{B}
\end{equation}%
accordingly to reactions (\ref{GrindEQ__50_}).
Timelike vector $\xi^i(x)$ defines the macroscopic and dynamic velocities of the system $v^i(x)$:
\begin{equation}\label{3.1.11}
v^i =
\frac{\xi^i}{\xi}\,; \quad (v,v) = 1\,,
\end{equation}
and its local temperature $\theta (x)$:
\begin{equation}\label{3.1.12}
\theta (x) = \xi^{-1},
\end{equation}
wherewith to define chemical potentials, $\mu_a(x)$, in the ordinary normalization:
\begin{equation}\label{3.1.13}
\mu_a(x) = \theta(x) \nu_a(x)\,.
\end{equation}
In these terms distribution \Req{3.1.9} can be written in following form:
\begin{equation}\label{3.1.14}
f^0_a(x,p_a) = \frac{1}{\displaystyle \mathrm{e}^
{\displaystyle \frac{- \mu_a + (v,p_a)}{\theta}} \mp 1},
\end{equation}
and moments of distribution take form \cite{Ignatev3},
\cite{kuza}:
\begin{eqnarray}\label{3.1.17} n^i_a(x) = n_a(x) v^i\,;\\
\label{3.1.18}
\stackunder{a}{T}^{ik}(x) = (\Eps_a + P_a) v^i v^k - P_a g^{ik},
\end{eqnarray}
where:\footnote{$\rho=2S+1$}
\newcommand{\fo}{\exp\bigl(\frac{- \mu_a + \sqrt{m^2_* + p^2}}{
\theta}\bigr)\mp 1}
\begin{equation}\label{3.1.19}
n_a(x) =
\frac{\rho}{2\pi^2} \displaystyle{\int\limits_{0}^{\infty}} \frac{ p^2 d p}{\fo}\,;
\end{equation}
\begin{equation}
\label{3.1.20}\Eps_a(x) =
\frac{\rho}{2\pi^2} \displaystyle{\int\limits_{0}^{\infty}} \frac{\sqrt{m^2_* + p^2} p^2 d p}{\fo}\,; \\
\end{equation}
\begin{eqnarray}
\label{3.1.21} P_a(x) =& \displaystyle{\frac{\rho}{6\pi^2}
\int\limits_{0}^{\infty}}\frac{p^4
d p}{\sqrt{m^2_* + p^2}}\nonumber \\
\times & \displaystyle\frac{1}{\fo}\,;
\end{eqnarray}
\begin{eqnarray}
\label{3.1.21_1}T_p= \Eps_a(x)-3P_a(x) =\displaystyle{\frac{\rho m_*^2}{2\pi^2}} \times  \nonumber \\
\displaystyle{\int\limits_{0}^{\infty}}\frac{p^2d p}{\sqrt{m^2_* + p^2}}
\displaystyle\frac{1}{\fo}\,;
\end{eqnarray}
\begin{eqnarray}
\label{3.1.21a} \sigma^{(r)}_a(x) = \displaystyle{\frac{\rho m_*q^{(r)}_a}{2\pi^2}
\displaystyle{\int\limits_{0}^{\infty}}} \frac{p^2
d p}{\sqrt{m^2_* + p^2}}\nonumber \\
\times \displaystyle\frac{1}{\fo} \equiv \frac{q^{(a)}_r}{m_*}T_p.
\end{eqnarray}
All macroscopic scalars at that are additive:
\begin{equation}\label{sum}
\Eps=\sum\limits_a \Eps_a; \; P=\sum\limits_a P_a; \; \sigma^{(r)}=\sum\limits_a \sigma^{(r)}_a.
\end{equation}
\subsection{Ultrarelativistic limit}
In the ultrarelativistic limit
\begin{equation}\label{ultralimit}
\frac{p}{m_*}\to \infty;\Rightarrow \mathcal{E}_{pl}-3P_{pl}\to 0;\quad \sigma\to 0
\end{equation}
the asymptotic expressions for macroscopic scalars
 (\ref{3.1.19}) -- (\ref{3.1.21a}) take form:
\newcommand{\fou}{\mathrm{e}^{\frac{-\mu+p}{\theta}}\mp 1}
\begin{equation}\label{3.1.19u}
n(x) =
\frac{\rho}{2\pi^2} \displaystyle{\int\limits_{0}^{\infty}} \frac{ p^2 d p}{\fou}=\theta^3\phi_2(\nu);
\end{equation}
\begin{equation}
\label{3.1.20u}\Eps_{pl} =
\frac{\rho}{2\pi^2} \displaystyle{\int\limits_{0}^{\infty}} \frac{p^3 d p}{\fou}=
\theta^4\phi_3(\nu);
\end{equation}
%
\begin{equation}
\label{3.1.22u} \sigma = \Phi \displaystyle{\frac{q^2\rho}{2\pi^2}
\int\limits_{0}^{\infty}}\frac{p d p}{\fou}=q^2\Phi\theta^2\phi_1(\nu),
\end{equation}
where we introduce the functions of reduced chemical potential $\nu=\mu/\theta$:
\begin{equation}\label{phi}
\phi_n(\nu)=\frac{\rho\nu^{n+1}}{2\pi^2} \displaystyle{\int\limits_{0}^{\infty}} \frac{ x^2 d x}{\mathrm{e}^{\nu(-1+x)}\mp 1}.
\end{equation}

\section{The Self - Consistent Kinetic Model of Self - Gravitating Plasma with Interpartial Scalar Interaction}
\subsection{The Lagrangian formalism}

In the article for the purpose of methodological simplicity we consider a system consisting of one scalar field, $\Phi$.
The generalization of results to the case of $n$ scalar fields with an account of above-cited formulas and additivity of the Lagrangian function does not require any specific efforts.

Let us consider the Lagrangian function
of classical massive real scalar field $\Phi$.
In such a case the Lagrangian scalar field can be chosen in the following form:
\begin{equation} \label{Ls} L_{s} =\frac{\epsilon_1}{8\pi } \left(g^{ik} \Phi _{,i} \Phi _{,k} -\epsilon_2 m_{s}^{2} \Phi ^{2} \right), \end{equation}
where $m_{s} $ is a mass of scalar field quanta,  $\epsilon_2=1$ for the classical scalar field,
$\epsilon_2=-1$ for a fantom (in terms of negativeness of the kinetic energy)
scalar field; $\epsilon_1=1$ for a field with repulsion of likely charged particles,
$\epsilon_1=-1$ for a field with attraction of likely charged particles $\epsilon_1=-1$.

Let us write down an invariant action function for the system of ``scalar charged particles + scalar field''  \cite{Yu_Sasha_Misha_Mitya}
 in order to get the Euler-Lagrange equation for a scalar field:
\begin{eqnarray}\label{S_pf}
S=S_p+S_s=\int\limits_\Omega d\Omega\times\nonumber\\
 \left(\sum\limits_a\int\limits_{-\infty}^{+\infty} m_*^{(a)}\delta^4(x,x_a(s_a))ds_a+L_s\right),
\end{eqnarray}
where $x_a(s_a)\equiv x^i_{a}(s_a)$ are parametric equations of particles' motion defined by their proper times $s_a$, $\Omega$ is a 4-d volume of Riemann space.
To get the Euler-Lagrange equation of the scalar field in accordance with a standard procedure it is necessary to calculate the variation $S$ at given particles' trajectory taking into account the formulas for the effective mass (\ref{m_*}) and Lagrange functions of the scalar field (\ref{Ls}) as well as arbitrariness of the scalar field variations.
 As a result of standard calculations we obtain:
\begin{equation}\label{EqF0}
\square\Phi+m_{s}^{2} \Phi =-4\pi \epsilon_1 \sigma,
\end{equation}
where
\[{\rm \square }\Phi \equiv g^{ik} \nabla _{i} \nabla _{k} \Phi =\frac{1}{\sqrt{-g} } \frac{\partial }{\partial x^{i} } \sqrt{-g} g^{ik} \frac{\partial }{\partial x^{k} } \Phi \]
is D'Alembert operator and scalar charge density
of system of particle moving along given trajectories
, $\sigma$, is described by formula
\begin{equation}
\sigma=\sum\limits_a q_a\int\limits_{-\infty}^{+\infty} \delta^4(x,x_a(s_a))ds_a. \nonumber
\end{equation}
The statistical averaging of the last expression with an account of properties of Dirac $\delta$ - function (methodic of similar calculations with usage of invariant functions of sources see e.g. in \cite{Bogolyub}) reduces it to form (\ref{3.1.21a}).

Let us now consider the Lagrangian function of classical massive real conformal scalar field $\Phi $\footnote{by conformal
invariance we understand here a asymptotic property attained in the ultrarelativistic limit $m_s\to0$}
 (see e.g. \cite{YuNewScalar3};
for massive scalar field the conformal
invariance is understood as asymptotic
property at ($m_{s} \to 0$)):
\begin{equation} \label{LsR} L_{s} =\frac{\epsilon_1}{8\pi } \left(g^{ik} \Phi _{,i} \Phi _{,k} +
\frac{R}{6}\Phi ^{2}   -\epsilon_2 m_{s}^{2} \Phi ^{2}\right). \end{equation}
The Lagrangian function differs from the standard one (see e.g. \cite{Melnikov})
for the presence of factor $1/8\pi$ and also for introduced unit indicators $\epsilon_\alpha$. Let us find the equation for scalar field using this function in the action integral
(\ref{S_pf}):
\begin{equation} \label{EqPhiR} {\rm \square }\Phi -\frac{R}{6} \Phi +\epsilon_2 m_{s}^{2} \Phi =
-4\pi \epsilon_1 \sigma.
\end{equation}
 Let us notice that in this paper we obtain Ricci tensor by convolution of first and third indices of Riemann tensor $R_{jl}=g^{ik}R_{ijkl}$.

\subsection{Energy-Momentum Tensor And the Conservation Laws}

Energy-momentum tensor of a scalar field relative to the Lagrangian function
(\ref{Ls}) is:
\begin{equation} \label{Tiks} T_{s}^{ik} =\frac{\epsilon_1}{8\pi } \left(2\Phi ^{,i} \Phi ^{,k} -
g^{ik} \Phi ^{,j} \Phi _{,j} +\epsilon_2 g^{ik} m_{s}^{2} \Phi ^{2} \right). \end{equation}
Einstein equations for the statistical system of scalar charged particles have from:
\begin{equation} \label{G_iks} R^{ik} -\frac{1}{2} Rg^{ik} =8\pi (T_{p}^{ik} +T_{s}^{ik} ), \end{equation}
where we need to substitute the expressions for the components of energy-momentum tensors of plasma
\eqref{3.1.18}, \eqref{3.1.20}, \eqref{3.1.21} and scalar field \eqref{Tiks}.
Calculating covariant divergences from both parts of Einstein equations \eqref{G_iks},
we obtain from \eqref{GrindEQ__57_} and \eqref{Tiks} the total energy-momentum conservation laws:
\begin{eqnarray} \label{Tikk} \nabla _{k} (T_{p}^{ik} +T_{s}^{ik} )=\nonumber\\
\frac{1}{4\pi }
\nabla ^{i} \Phi \left[\epsilon_1({\rm \square }\Phi +\epsilon_2 m_{s}^{2} \Phi )+4\pi \sigma \right]=0,
\end{eqnarray}
where from, putting $\Phi \not\equiv {\rm Const}$, we again obtain the equation for a massive non-conformal scalar field with a source (\ref{EqF0}) \cite{Ignatev3}.

The trace of energy-momentum tensor of the scalar field \eqref{Tiks} is equal to:
\begin{equation} \label{SpTik}
T_{s} =\frac{\epsilon_1}{4\pi } (-\Phi ^{,j} \Phi _{,j} +2\epsilon_2 m_{s}^{2} \Phi ^{2} ).
\end{equation}

The components of scalar field's energy-momentum tensor relative to the Lagrangian function
 \eqref{LsR} are \cite{Melnikov}:
\begin{equation} \label{T_iksR}
\begin{array}{l}
T_{s}^{ik} =\displaystyle{\frac{\epsilon_1}{4\pi }} \biggl[\Phi^{,i}\Phi ^{,k} -\frac{1}{2}g^{ik}\Phi_{,j}\Phi^{,j} + \frac{1}{2}\epsilon_2 m_{s}^{2} g^{ik} \Phi ^{2} + \\
\frac{1}{6} \biggl(R^{ik} -\frac{1}{2} Rg^{ik} \biggr)\Phi ^{2} -\frac{1}{6} \bigl(\nabla^i\nabla^k -g^{ik}\Box\bigr)\Phi^2 \biggr].
\end{array}
\end{equation}
Carrying out differentiation of $\Phi^2$ in this expression, we get another notation of the scalar field's energy momentum tensor (see e.g. \cite{Yubook1})
\begin{equation} \label{T_iksR1}
\begin{array}{l}
T_{s}^{ik}=\displaystyle{\frac{\epsilon_1}{8\pi }} \biggl[\frac{4}{3}\Phi^{,i}\Phi ^{,k}-\frac{1}{3}g^{ik}\Phi_{,j}\Phi^{,j}+\epsilon_2 m^2_sg^{ik}\Phi^2+\\
\frac{1}{3} \biggl(R^{ik} -\frac{1}{2} Rg^{ik} \biggr)\Phi ^{2}-\frac{2}{3}\Phi\Phi^{,ik}+\frac{2}{3}g^{ik}\Phi\Box\Phi\biggr].
\end{array}
\end{equation}
The covariant divergence of tensor \eqref{T_iksR1} with an account of commutation relations
for second covariant derivatives of vector (see e.g. \cite{Petrov}):
\begin{eqnarray}\label{[R]}
u_{i,kl}-u_{i,lk}=R^m_{.ikl}u_m\Rightarrow \nonumber\\
g^{kl}\nabla_k\bigl(\nabla_l\Phi_{,i})=\nabla_i \Box\Phi+R^k_i\nabla_k\Phi \nonumber
\end{eqnarray}
is:
\begin{equation} \label{T_iksRk} \nabla _{k} T_{s}^{ik} =\frac{\epsilon_1}{4\pi }
\nabla ^{i} \Phi \left(\Box\Phi -
\frac{R}{6} \Phi+\epsilon_2 m_{s}^{2} \Phi  \right), \end{equation}
Calculating covariant divergences from both parts of Einstein equations \eqref{G_iks},
we obtain from \eqref{GrindEQ__57_} and \eqref{T_iksRk} the total energy-momentum conservation laws:
\begin{eqnarray} \label{TikkR}
\nabla _{k} (T_{p}^{ik} +T_{s}^{ik} )=\frac{1}{4\pi }\nabla ^{i} \Phi \times\nonumber\\
\left[\epsilon_1 \left({\rm \square }\Phi-
\frac{R}{6} \Phi +\epsilon_2 m_{s}^{2} \Phi  \right)+4\pi \sigma \right]=0,
\end{eqnarray}
where from, putting $\Phi \not\equiv {\rm Const}$, we obtain the equation (\ref{EqPhiR}) for a massive scalar field with a source.

Calculating the trace of the scalar field's energy-momentum tensor (\ref{T_iksR1}), we find:
\begin{equation}\label{Ts}
T_s\equiv g_ikT^{ik}_s=\frac{\epsilon_1}{4\pi}\Phi\biggl(\Box\Phi -
\frac{R}{6} \Phi+2\epsilon_2 m_{s}^{2} \Phi  \biggr),
\end{equation}
where from, with an account of field equation (\ref{EqPhiR}) we get a simplified expression:
\begin{equation} \label{TsR}
T_{s} =\frac{\epsilon_1\epsilon_2}{4\pi } m_{s}^{2} \Phi ^{2} - \sigma \Phi.
\end{equation}
%
\subsection{The Complete System of Equations of the Kinetic Model of Self-Gravitating System of Scalarwise Interacting Particles
at Local Thermodynamic Equilibrium Conditions}
The complete system of equations of the kinetic model of self-gravitating system of scalarwise interacting particles at conditions of local thermodynamic equilibrium includes Einstein equations (\ref{G_iks}), transport equations of energy-momentum of particles (\ref{GrindEQ__57_}), scalar charge conservation law (\ref{GrindEQ__55_}) (if the charge is conserved), equations of chemical equilibrium (\ref{lambda_chem}), equations of scalar field(s) (\ref{EqF0}) or (\ref{EqPhiR}) together with definitions of current vector (\ref{3.1.17}),
statistical system's energy-momentum tensor (\ref{3.1.18}), macroscopic scalars (\ref{3.1.19}) -- (\ref{3.1.21a}),
and energy-momentum tensor of scalar field (\ref{Tiks}) or (\ref{T_iksR1}).

It is appropriate to make here a following notice. The conservation laws of the total energy-momentum tensor of system ``particles + scalar fields''
(\ref{Tikk}) or (\ref{T_iksRk}) are identically fulfilled at constant scalar fields
\begin{eqnarray}\label{Phi0}
\nabla_i\Phi=0\Leftrightarrow \Phi=\Phi_0=\mathrm{Const}\\
\Rightarrow \nabla _{k} (T_{p}^{ik} +T_{s}^{ik} )\equiv0.\nonumber
\end{eqnarray}
However, imposing condition (\ref{Phi0}) on scalar fields contradicts the principle of least action for such fields with this principle being a fundamental principle of theoretical physics.
 This contradiction is valid for physical scalar fields.
However if such ``scalar fields'' are generated by the cosmological constant or theories of gravitation of type $f(R)$ such condition is legitimate but such fields are not physical and of geometric origin. In this article we exactly consider physical scalar fields with sources.

\subsection{Generation of Mass of Conformal Invariant Scalar Field with a Source}
Let us consider equation of conformal-invariant scalar field (\ref{EqPhiR}) in case of locally equilibrium system of ultrarelativistic scalar chaged particles. Using expression for scalar charge density (\ref{3.1.22u})
we reduce this equation to form:
\begin{equation} \label{EqPhiR} {\rm \square }\Phi -\frac{R}{6} \Phi+(\epsilon_2 m^2_s+\epsilon_1 m^2_{ef})\Phi  =0,
\end{equation}
where the following denotation is introduced:
\begin{equation}\label{m_eff}
m^2_{ef}=4\pi q^2\theta^2\phi_1(\nu).
\end{equation}
\begin{thm}\label{stat1}
1. Scalar charge density $\sigma$ for ultrarelativistic locally equilibrium system of particles plays the role of scalar field's effective mass, which, generally speaking, depends on coordinates.\\
2. Indicator $\epsilon_1$ at that plays the same role as indicator $\epsilon_2$ for standard massive member in the equation of scalar field: case of likely charged particles repulsion corresponds to value $\epsilon_1=+1$, case of likely charged particles attraction corresponds to  $\epsilon_1=-1$.\\

3. At $\epsilon_1\epsilon_2=-1$ there exists a hypothetical possibility of complete nulling of summary massive member
$m^2_s-m^2_{ef}=0$ and strict recovery of conformal invariance of the scalar field equation.
\end{thm}

\section{Self-Consistent Cosmological Model for Local Equilibrium Plasma with Interpartial Scalar Interaction}
\subsection{A Model of Local Thermodynamic Equilibrium}
If local equilibrium condition is fulfilled (LTE) then
\begin{equation}\label{LTE}
t\gg \tau_{ef},
\end{equation}
(where $t$ is a characteristic time scale of the statistical systems, $\tau_{ef}$ is an effective time of interparticle interactions) integral of collisions in the right part of the kinetic equations becomes a greater value therefore for local equilibrium plasma it is necessary to use the definition of energy-momentum tensor of liquid \eqref{3.1.18} and relations \eqref{3.1.19} -- \eqref{3.1.21}  defining macroscopic scalars and equations of chemical equilibrium \eqref{lambda_chem} rather than kinetic equations solution. It should be considered that if chemical equilibrium conditions are fulfilled, locally equilibrium distribution functions \eqref{3.1.9} automatically make integral of collisions \eqref{GrindEQ__53_} equal to zero. However, according to logics of hydrodynamic approximation (see e.g. [1]) equality to zero of the right part of the kinetic equations in this case should be seen as just an approximate relation valid only for the macroscopic moments of the distribution function. Let us notice the following important circumstance. As is known, formally in LTE case matter equations obtained on the basis of the kinetic theory do not differ from the equations of hydrodynamics. As is known, formally in LTE case matter equations obtained on the basis of the kinetic theory do not differ from the hydrodynamic equations. However it is also known that hydrodynamic equations do not represent a closed system of equations. To make this system closed it is required to add relations of coupling between macroscopic scalars. This is not required at kinetic approach since corresponding functional relations are actually contained in the integral definitions of macroscopic scalars which does not allow any space for speculations.

Further, let us notice that the problem of LTE establishment with respect to specific interactions in the expanding Universe is quite a delicate question and requires a specific research.%

 Condition (\ref{LTE}), as it turns out, is defined through the dependency of total cross-section of particles' interaction on the first kinematic invariant $s=(p_a+p_b)^2$ of particles' pair interactions. If approximate the value of total cross-section of particles' interaction in range of high energies by power dependence  $\sigma_{tot}\backsim s^\alpha$, then for ultrarelativistic particles in the early Universe in case of summary barotropic equation of state $P=\kappa\mathcal{E}$ condition (\ref{LTE}) brings us to the following statement \cite{LTE}:\footnote{it is assumed at that that particles' concentration is defined by locally-equilibrium formula (\ref{3.1.19})}
\begin{thm}\label{LTEthm}
1. At $\kappa\not=-1$ and the next condition's fulfillment
\begin{equation}\label{Yus_8}
\alpha > -\frac{3}{4}(1-\kappa)
\end{equation}
LTE is maintained at early expansion stages and violated at the late ones whereas at fulfillment of the condition invert to (\ref{Yus_8}) LTE is violated at the early stages and recovered on the late ones. \\

2. In the inflation case ($\kappa=-1$) at:
\begin{equation}\label{Yus_9}
\alpha >-\frac{3}{2}
\end{equation}
LTE is maintained on the early stages and violated on the late ones.
\end{thm}

\subsection{The Self-Consistent System of Equations for the Isotropic Homogenous Space-flat Universe}

Let us consider the space-flat Friedmann cosmological model
\begin{eqnarray} \label{ds2} ds^{2} =a^2(\eta)d\bar{s}^2\equiv \nonumber\\
a^2(\eta)(d\eta^{2} -dx^{2} -dy^{2} -dz^{2} ),
\end{eqnarray}
where matter comprises of equilibrium plasma of scalarwise interacting particles and massive scalar field, depending only on cosmological time $\Phi(t)$. Состоянию покоя плазмы относительно синхронной в метрике \eqref{ds2} системы отсчета соответствует вектор макроскопической скорости:
\begin{equation} \label{vi} v^{i} =a^{-1}\delta _{4}^{i} . \end{equation}
Einstein metric's components relative to metrics \eqref{ds2} are equal to:
\begin{equation} \label{Gik_Fr}
G^{i}_{k} =2\frac{2a'^2-aa''}{a^4} v^{i} v_{k} +\frac{2aa''-a'^2}{a^4}\delta _{k}^{i}.
\end{equation}
Further, calculating components $\Phi _{\; ,k}^{,i} $, we find:
\begin{equation} \label{Phiik}
\Phi _{\; ,k}^{,i} =\biggl(\Phi''-2\frac{a'}{a} \Phi'\biggr)\frac{v^{i} v_{k}}{a^2} +\frac{a'\Phi'}{a^3} \delta _{k}^{i}.
\end{equation}
From \eqref{Gik_Fr} and \eqref{TsR} it follows:
\begin{equation} \label{GrindEQ__91_} -R=6\frac{a''}{a^{3} } =-8\pi\sigma\Phi +2\epsilon_1\epsilon_2 m_{s}^{2} \Phi ^{2}.
\end{equation}
Taking into account relations \eqref{Gik_Fr}--\eqref{GrindEQ__91_},
we can calculate components of energy-momentum tensor of a scalar field
and represent them in form of
components of tensor of the ideal flux
\begin{equation} \label{GrindEQ__93_} \mathop{T}\limits_{s} {\rm \; }_{k}^{i} =(\mathcal{E}_{s} +P_{s} )v^{i} v_{k} -P_{s} \delta _{k}^{i} , \end{equation}

Herewith for non-conformal invariant scalar field
we find from \eqref{Tiks}:
\begin{equation} \label{ESR}
\mathcal{E}_{s} =\frac{\epsilon_1}{8\pi} \biggl(\frac{\Phi'^2}{a^2}+m_{s}^{2}\Phi^{2}\biggr);
\quad P_{s} =\frac{\epsilon_1}{8\pi}\biggl(\frac{\Phi'^2}{a^2}-m_{s}^{2}\Phi^{2}\biggr). \end{equation}
Moreover, we obtain the following relation:
\begin{equation} \label{GrindEQ__97_}
T_{s} =\mathcal{E}_{s} -3P_{s} =\frac{\epsilon_1}{4\pi } (-\frac{\Phi'^2}{a^2} +2m_{s}^{2} \Phi ^{2} ),
\end{equation}
and:
\begin{equation}
\label{GrindEQ__98_}
\mathcal{E}_{s} +P_{s} =\frac{\epsilon_1}{4\pi}\frac{\Phi'^2}{a^2}.
\end{equation}
Herewith scalar field equation (\ref{EqF0}) takes form:
\begin{equation}\label{EqF0}
\frac{1}{a^4}\frac{d}{d\eta}a^2\frac{d}{d\eta}\Phi+m_{s}^{2} \Phi =-4\pi \epsilon_1 \sigma.
\end{equation}
The mathematical model of cosmological evolution descried above in case of conformal non-invariant scalar field has been researched in details in the series of articles cited above an, in particular, in \cite{Yu_Sasha_Misha_Mitya}.
Models of such kind ensure Universe acceleration and allow the possibility of anomalously fast expansion.

\section{Conformal Transformations of the Cosmological Model Equations}

In this article we consider conformal-invariant models of scalar field.
In case of conformal-invariant scalar field let us notice the following relation useful in the future:
\begin{equation}\label{f/a}
\biggl(\Box-\frac{R}{6} \biggr)\frac{\phi}{a}=\frac{\phi''}{a^3},
\end{equation}
using which the equation of conformal-invariant scalar field \eqref{EqPhiR} $\Phi(\eta)$
in metrics \eqref{ds2} can be written in the following form:
\begin{equation} \label{EqS_Freed}
\frac{1}{a^3}\frac{d^2}{d\eta^2}a\Phi+\epsilon_2 m^2_s\Phi=-4\pi\epsilon_1\sigma.
\end{equation}
Taking into account relations \eqref{Gik_Fr}--\eqref{GrindEQ__91_},
let us calculate the components of energy-momentum tensor of conformally scalar field \eqref{T_iksR} and fine:
\begin{equation} \label{GrindEQ__94_}
\mathcal{E}_{s} =\frac{\epsilon_1}{8\pi } \left[\frac{1}{a^4} (a\Phi)'^{2} +\epsilon_2 m_{s}^{2} \Phi ^{2} \right]; \end{equation}
\begin{equation} \label{GrindEQ__95_} P_{s} =\frac{\epsilon_1}{24\pi } \left[\frac{1}{a^4} (a\Phi)'^{2} -\epsilon_2 m_{s}^{2} \Phi ^{2}
+8\pi\epsilon_1\Phi\sigma\right]. \end{equation}

\subsection{Conformal Transformations of Macroscopic Scalars}
Let us investigate the transformational properties of macroscopic scalars (\ref{3.1.19}) -- (\ref{3.1.21a})
with respect to conformal transformations:
\begin{equation}\label{conf_trans}
ds^2=a^2(\eta)d\bar{s}^2;\quad \Phi(\eta)=\frac{\bar{\Phi}(\eta)}{a(\eta)}.
\end{equation}
At transformation of momentum and thermodynamic scalars by the next law:
\begin{equation}\label{term_conf}
p=\frac{\bar{p}}{a(\eta)};\quad \theta=\frac{\bar{\theta}}{a(\eta)}; \quad \mu=\frac{\bar{\mu}}{a(\eta)}
\end{equation}
the macroscopic scalars (\ref{3.1.19}) -- (\ref{3.1.21a}) are transformed by laws:
\begin{eqnarray}\label{macr_conf}
n_a=\frac{\bar{n}_a}{a^3(\eta)}; & \displaystyle\mathcal{E}_a=\frac{\bar{\mathcal{E}}_a}{a^4(\eta)};\nonumber\\
P_a=\frac{\bar{P}_a}{a^4(\eta)}; & \displaystyle\sigma_a=\frac{\bar{\sigma}_a}{a^3(\eta)}.
\end{eqnarray}
Then the fundamental current $Q$ conservation law takes form:
\begin{equation}\label{dn=0}
\frac{1}{a^4}\frac{d}{d\eta}a^3\sum\limits_a q_a n_a=0\Rightarrow \sum\limits_a q_a \bar{n}_a=\mathrm{Const}.
\end{equation}

The transport equation of energy-momentum of particles \eqref{GrindEQ__57_} in metrics \eqref{ds2} can be written in the following form:
\begin{equation} \label{GrindEQ__101_} \mathcal{E}'_{pl} +3\frac{a'}{a} (\mathcal{E}_{pl} +P_{pl} )=\sigma \Phi' . \end{equation}
Then, taking into account relation (\ref{3.1.21a}), we find:
\begin{equation}
\label{3.1.21au} \sigma = \frac{\Eps_{pl}-3P_{pl}}{\Phi}.
\end{equation}
Carrying out conformal transformations (\ref{conf_trans}) -- (\ref{term_conf})  in equation (\ref{GrindEQ__101_} with an account of (\ref{macr_conf})
and (\ref{3.1.21au}) we find:
\begin{equation}\label{conf_E'}
\bar{\mathcal{E}}'_{pl}-(\bar{\Eps}_{pl}-3\bar{P}_{pl})\frac{\bar{\Phi}'}{\bar{\Phi}}=0.
\end{equation}
In the ultrarelativistic limit
\begin{equation}\label{ultralimit}
\frac{p}{m_*}\to \infty;\Rightarrow \bar{\mathcal{E}}_{pl}-3\bar{P}_{pl}\to 0;\quad \bar{\sigma}\to 0
\end{equation}
equation (\ref{conf_E'}) is reduced to
\begin{equation}\label{conf_E_const}
\bar{\mathcal{E}}'_{pl}=0\Rightarrow \bar{\mathcal{E}}=\bar{\mathcal{E}}_0=\mathrm{Const}\Rightarrow \mathcal{E}_a=\frac{\mathrm{Const}}{a^4}.
\end{equation}
From the conservation law of scalar charged particles
\begin{equation}\label{n_const}
\bar{n}=\bar{\theta}^3\gamma^3\phi_2(\gamma)=\mathrm{Const},
\end{equation}
and also from (\ref{conf_E_const}) we find:
\begin{equation}\label{E_const}
\bar{\mathcal{E}}_{pl}=\bar{\theta}^4\gamma^4\phi_3(\gamma)=\mathrm{Const}.
\end{equation}
Thus, we have two functionally independent equations (\ref{n_const}) and  (\ref{E_const}) on two functions:
$\bar{\gamma}(\eta)$ and $\bar{\theta}(\eta)$. Arbitrary constants can be the unique solution of these equations:
\begin{eqnarray}\label{theta_gamma}
\bar{\gamma}=\mathrm{Const};& \bar{\theta}=\mathrm{Const}\Rightarrow\nonumber\\
\theta=\frac{\bar{\theta}_0}{a(\eta)};& \displaystyle \mu=\frac{\bar{\mu}_0}{a(\eta)},
\end{eqnarray}
which ensures conformal invariance of matter equations in the ultrarelativistic limit.

\subsection{Conformal Transformations of Scalar Field Equation}
Let us turn our attention to the field equations. At zero massive member in equation of scalar field
 (\ref{EqS_Freed}) we get field equation's transformation law:
\begin{equation}\label{EqS_Freed_Conf}
\frac{1}{a^3}\frac{d^2}{d\eta^2}a\Phi=-4\pi\epsilon_1\sigma\Rightarrow \frac{d^2}{d\eta^2}\bar{\Phi}=-4\pi\epsilon_1\bar{\sigma},
\end{equation}
where
\begin{equation}\label{bar_sigma}
\bar{\sigma}= q^2\bar{\theta}^2\gamma^2\phi_1(\gamma)\bar\Phi\equiv \frac{\omega^2_0}{4\pi }\bar{\Phi},
\end{equation}
where
\begin{equation}\label{omega0}
\omega_0=|q|\bar{\theta}\gamma\sqrt{4\pi\phi_1(\gamma)}=\mathrm{Const}.
\end{equation}

Thus, in the ultrarelativistic limit $\bar{\sigma}$ depends on $\eta$ only be means of $\bar{\Phi}(\eta)$ and equation
 (\ref{EqS_Freed_Conf}) has its solution in the ultrarelativistic limit:
\begin{eqnarray}\label{cos}
\bar{\Phi}=& C_1\cos\omega_0\eta+C_2\sin\omega_0\eta;& \epsilon_1=+1;\\
\label{exp}
\bar{\Phi}=& C_1\mathrm{e}^{\omega_0\eta}+C_2\mathrm{e}^{-\omega_0\eta};& \epsilon_1=-1,
\end{eqnarray}

Let us notice that in the ultrarelativistic limit (\ref{ultralimit}) $\omega_0\to0$, thus according to (\ref{cos}) and (\ref{exp}) in this approach we can put
:
\begin{equation}\label{omega_eta}
\omega_0\eta\to 0,
\end{equation}
hence in the ultrarelativistic limit (\ref{ultralimit}) an asymptotic solution takes place:

\begin{equation}\label{}
\bar{\Phi}\backsimeq \bar{\Phi}_0=\mathrm{Const}\Rightarrow \Phi\backsimeq \frac{\bar{\Phi}_0}{a(\eta)},
\end{equation}
i.e. the asymptotic conformal invariance is recovered also for scalar field.

Let us notice that in this approximation in accordance with (\ref{GrindEQ__94_})
scalar field's energy density and pressure are equal to zero $\mathcal{E}_s=0$, and the unique non-trivial solution of
Einstein equation:
\begin{equation} \label{GrindEQ__99_} 3\frac{a'^{2} }{a^{4} } =8\pi \mathcal{E}_{pl}; \end{equation}
takes form:
\begin{equation} \label{Einst_Ultra_Conf} 3\frac{a'^{2} }{a^{4} } =8\pi \frac{\bar{\mathcal{E}}_{pl}}{a^4}\Rightarrow ; a'=\mathrm{Const}, \end{equation}
i.e. has the ultrarelativistic solution $a=a_1\eta$.

Thus we can make the following statement:

\begin{thm}\label{stat2}
1. For massless conformal-invariant scalar field in the ultrarelativistic limit(\ref{ultralimit}) and (\ref{omega_eta}) there is an asymptotically exact solution of Einstein equations
\begin{eqnarray}\label{exact}
a(\eta)=a_0\eta & \displaystyle \Phi=\frac{\Phi_0}{\eta} & \displaystyle \theta=\frac{\theta_0}{\eta};\nonumber \\
n=\frac{n_0}{\eta^3} & \displaystyle \sigma=\frac{\sigma_0}{\eta^3} & \displaystyle \mathcal{E}_{pl}=\frac{\mathcal{E}^0_{pl}}{\eta^4}.
\end{eqnarray}
2. Herewith accorging to  (\ref{m_*}) and (\ref{theta_gamma}) the factor of ultrarelativity remains constant:
\begin{equation}\label{p/m}
\frac{<p>}{m_*}\backsim \mathrm{max}\bigl(\frac{\theta}{m_*},\frac{\mu}{m_*} \bigr)=\mathrm{Const}.
\end{equation}
\end{thm}
\section{Discussion of the Results}
Summing up results of the research of the mathematical model of statistical system with interparticle conformal-invariant scalar interaction constructed on the basis of common-relativistic theory and gravitation theory, let us notice the most important properties of this model.
\begin{enumerate}
\item In case of ultrarelativistic locally equilibrium statistical system its scalar charge density $\sigma$ plays a role of massive member in conformal-invariant scalar field equation.
In particular, according to (\ref{m_eff}) in case of homogenous isotropic Universe
quanta effective mass of scalar field
is inversely proportional to
scale factor:
\begin{equation}\label{m_eff(eta)}
m_{ef}= \frac{\omega_0}{a(\eta)},
\end{equation}
i.e. is changed by the same law as effective mass of scalar charged particles provided (\ref{omega_eta}).
\item Herewith ultrarelativism factor $<p>/m_*$ remains constant and conserves its great value.
\item At the same conditions the entire Universe evolves by ultrarelativistic
law.
\end{enumerate}
As is well-known, in completely invariant theories there is no fixed scale. Apparently, because of this property conformal-invariant theories were thrown away in due time by cosmologists. However, as follows from the previous results, such a typical scale appears in such theories because of a scalar charge. This typical scale of mass is equal to $\omega_0$ and appears at violation of condition
(\ref{omega_eta}) simultaneously with violation of conformal invariance of the theory.

\section{Acknowledgements}

The author is grateful to participants of the seminar for relativistic kinetics and
cosmology (MW) of the Kazan Federal University for helpful
discussion of the work.


\begin{thebibliography}{}
%
%
\bibitem{Ignatev1}
Yu.G. Ignat'ev, Russ. Phys. J., 25, 372-375 (1982).
\bibitem{Ignatev2}
Yu.G. Ignat'ev, Russ. Phys. J., 26, 686-690 (1983).
\bibitem{Ignatev3}
Yu.G. Ignat'ev, Russ. Phys. J., 26, 690-694  (1983).
\bibitem{Ignatev4}
Yu.G. Ignat'ev, Russ. Phys. J., 26, 1068-1072 (1983).
%
\bibitem{kuza}%
Yu. G. Ignat'ev, R.R. Kuzeev, Ukr. Phys. J, 29, 1021 -- 1026 (1984).
%
\bibitem{YuNewScalar1}
Yu. G. Ignat'ev, Russ. Phys. J., 55, 166-172 (2012); 
arXiv:1307.1787v1 [gr-qc].%
\bibitem{YuNewScalar2}
Yu. G. Ignatiev (Ignat'ev), Russ. Phys. J., 55,  550-560, (2012); 
arXiv:1307.2472 [gr-qc]. %
\bibitem{YuNewScalar3}
Yu. G. Ignatyev (Ignat'ev), Russ. Phys. J., 55, 1345-1350 (2013); 
arXiv:1307.2509 [gr-qc].
%
\bibitem{YuNewScalar4}
Yu.G. Ignatyev (Ignat'ev) and D.Yu. Ignatyev, Grav. and Cosmol.,  Vol. 20, No. 4, pp. 299–303 (2014); arXiv:1408.3404v1 [gr-qc].
%
\bibitem{YuNewScalar5}
Yu.G. Ignatyev (Ignat'ev), A.A. Agathonov and D.Yu. Ignatyev,  Grav. and Cosmol., Vol.  20, pp. 304-308 (2014); arXiv:1408.3419v1 [gr-qc].
%
\bibitem{Yu_stfi14}
Yu. G. Ignat'ev, Space, Time and Foudamental Interections, No 1. - pp. 47-69 (2014).
%
\bibitem{Yu_stfi15}
Yu.G. Ignat'ev,	Space, Time and Foudamental Interections. – No 1 – pp. 5-23 (2015).
%
\bibitem{YuNewScalar6}
Yu. G. Ignatyev (Ignat'ev), Grav. and Cosmol., Vol. 21, No. 2, pp. 113–117 (2015); arXiv:1410.2487v1 [physics.ge].
%
\bibitem{Ignatev_print1}
Yu.G. Ignatyev (Ignat'ev), Grav. and Cosmol., (in print); arXiv:1504.02768v1 [gr-qc].
\bibitem{Ignatev_print2}
Yu.G. Ignatyev (Ignat'ev), Grav. and Cosmol., (in print);	arXiv:1504.03649v1 [gr-qc].
%
\bibitem{Yubook1}
Yurii G. Ignatyev (Ignat'ev). Relativistic Kinetic Theory of Nonequilibrium Processes in Gravitational
Fields. Kazan, Foliant-Press, -- 2010;  http://rgs.vniims.ru/books/const.pdf.
%
\bibitem{Yubook2}
Yurii G. Ignatyev (Ignat'ev). The Nonequilibrium Universe: The Kinetics Models of the Cosmological Evolution, Kazan: Kazan University Press, 2013;
http://www.stfi.ru/archive\_rus/2013\_2\_ Ignatiev.pdf
%
\bibitem{Mif}
Yu.G. Ignatyev and R.F. Miftakhov, Grav. and Cosmol. -- 2011, Vol. 17, No. 2, pp. 190–193; arXiv:1011.5774[gr-qc].
\bibitem{Yu_Sasha}
Yu. G. Ignatyev (Ignat'ev) and A. A. Agathonov,	Grav. and Cosmol. -- 2015, Vol. 21, No. 2, pp. 105–112; arXiv:1408.4738v1 [gr-qc].
\bibitem{Yu_Sasha_Misha_Mitya_stfi}
Yurii Ignat’ev, Alexander Agathonov, Mikhail Mikhailov and Dmitry Ignatyev,	 Space, Time and Foudamental Interections, No 3, pp. 16-31 (2014).
\bibitem{Yu_Sasha_Misha_Mitya}
Yurii Ignat’ev, Alexander Agathonov, Mikhail Mikhailov and Dmitry Ignatyev, Astrophys. Space Sci., 357:61 (2015); arXiv:1411.6244v1 [gr-qc].
\bibitem{Yu_Misha}
Yu. G. Ignat’ev, M. L. Mikhailov, Russ. Phys. J.  57, pp. 1743-1752 (2015).
%
\bibitem{Bogolyub}
Yu.G. Ignat'ev, Grav. and Cosmol.,  13, pp. 59-81 (2007).
%
\bibitem{Cartan}
E. Cartan, Les espaces de Finsler, Paris, 1934.
%
\bibitem{Vlasov}
A.A. Vlasov. Statistical Distribution Functions. Moskow, Nauka, 1966.
%
\bibitem{Landau_Stat}
L.D. Landau, E.M. Lifshitz. Statistical Physics. Vol. 5 (3rd ed.). Pergamon Press. %
Oxford$\cdot$ New York$\cdot$ Toronto$\cdot$ Sydney$\cdot$ Paris$\cdot$ Frankfurt, 1980.
%
\bibitem{Petrov}
A.Z. Petrov, Einstein spaces. Published by Pergamon Press   (1969).
%
\bibitem{Melnikov}
Vitaly N. Melnikov, Fields and Constants in the Theory of Gravitation, CBPF-MO-002/02, Rio de Janeiro, 2002.
%
\bibitem{LTE}
Yu.G. Ignat'ev,	Grav. and Cosmol.,  13 No 1, p. 31—42 (2007).
\end{thebibliography}
\end{document}